\documentclass{article}
\usepackage{graphicx}
\usepackage{color}
\usepackage{multirow}
\usepackage{comment}
\usepackage{url}
\usepackage{latexsym}
\usepackage[fleqn]{amsmath}
\usepackage{dcolumn}
\usepackage{siunitx}

\title{ Impact of arbitrage between leveraged ETF and futures on market liquidity during market crash}

\providecommand{\keywords}[1]
{
  \small	
  \textbf{\textit{Keywords---}} #1
}

\author{%
Ryuki Hayase$^{1}$, Takanobu Mizuta$^{2}$, Isao Yagi$^{3}$  \\
        \small $^{1}$Department of Informatics, Graduate School of Engineering, Kogakuin University \\
        \small $^{2}$SPARX Asset Management Co., Ltd. \\
\small $^{3}$Department of Information Science, Faculty of Informatics, Kogakuin University
}

\begin{document}

\maketitle
\begin{abstract} 
Leveraged ETFs (L-ETFs) are exchange-traded funds that achieve price movements several times greater than an index by holding index-linked futures such as Nikkei Stock Average Index futures.
It is known that when the price of an L-ETF falls, the L-ETF uses the liquidity of futures to limit the decline through arbitrage trading. Conversely, when the price of a futures contract falls, the futures contract uses the liquidity of the L-ETF to limit its decline.
However, the impact of arbitrage trading on the liquidity of these markets has been little studied. Therefore, the present study used artificial market simulations to investigate how the liquidity (Volume, SellDepth, BuyDepth, Tightness) of both markets changes when prices plummet in either (i.e., the L-ETF or futures market), depending on the presence or absence of arbitrage trading.
As a result, it was found that when erroneous orders occur in the L-ETF market, the existence of arbitrage trading causes liquidity to be supplied from the futures market to the L-ETF market in terms of SellDepth and Tightness.
When erroneous orders occur in the futures market, the existence of arbitrage trading causes liquidity to be supplied from the L-ETF market to the futures market in terms of SellDepth and Tightness, and liquidity to be supplied from the futures market to the L-ETF market in terms of Volume.
We also analyzed the internal market mechanisms that led to these results.
\end{abstract}


\begingroup
\renewcommand\thefootnote{}
\footnotetext{ The present paper is an English translation of the authors' original paper published in the\textit{The IEICE Transactions on Information and Systems} (Vol. 109-D, No. 1, 2026). The original Japanese version is available at: DOI: 10.14923/transinfj.2025SGP0003. \copyright copyright 2006 IEICE}
\addtocounter{footnote}{-1}
\endgroup

% 本文
\section{Introduction}
%\section{まえがき}
Leveraged ETFs are exchange-traded funds that achieve price movements several times greater than the underlying index by holding index-linked futures such as Nikkei Stock Average futures.
\par For example, a leveraged ETF with 2x leverage will move twice as much as the index it tracks.
Since these leveraged ETFs are listed, they can be traded on an exchange.
 When a leveraged ETF has 2x leverage, it holds a large quantity of futures contracts on the underlying index to achieve twice the price movement of that index.
Additionally, to maintain leverage, the holdings of the underlying index must be adjusted.
To adjust holdings, if the index rises (falls), it is necessary to buy (sell) futures.
This buying and selling to adjust holdings is called rebalancing.
While rebalancing does not have a strict schedule, it is often performed at the end of the day. It is said that this rebalancing can potentially amplify the rise (fall) of the underlying index further\cite{BloombergMay28th2014}.
Empirical studies have examined the impact of leveraged ETFs on their underlying assets\cite{BBSH12,SHHR16,Tr10}. In this regard, Qing et al.\cite{BBSH12} found that the returns of constituent stocks experiencing increased rebalancing transactions tend to be higher.
As for the impact on volatility, Shum et al.\cite{SHHR16} found that rebalancing transactions affect volatility at market close.
Trainor Jr.\cite{Tr10} examined the impact of leveraged ETF rebalancing transactions on volatility and reported that although rebalancing does not affect volatility in large underlying markets, it may have an impact in smaller markets.
Research on the underlying assets of leveraged ETFs has also included studies comparing the long-term performance of leveraged ETFs with their underlying assets\cite{AZ10,LWZ09}.
Recent studies on leveraged ETFs include ones on rebalancing transaction models incorporating market friction\cite{DKSY21} and on the log returns of leveraged ETFs over long periods\cite{B24}.

\par When investigating the impact of arbitrage trading between two markets, like in the present study, on market liquidity in real financial markets through empirical research, it is difficult to isolate and analyze the effects of arbitrage trading alone due to the involvement of various factors.
As a method to resolve such challenging issues, simulations using artificial markets have been conducted,
where an artificial market is an agent simulation system that constructs a virtual financial market on a computer\cite{MY25}.

\par Artificial markets allow for the isolation and analysis of specific influences, thereby enabling investigations into such effects as the impact of tick size on market liquidity\cite{YZY20}, as well as the effects of price limits\cite{ZPZLX16}.
Research has been reported not only on stock markets but also on cryptocurrency markets\cite{CCM17,CTM19}. Furthermore, some studies using artificial markets have investigated the impact of arbitrage trading on market liquidity in ETF markets\cite{GMY24}.
The present study examined representative market liquidity metrics-Volume, Depth, and Tightness-in leveraged ETFs versus non-leveraged ETFs, comparing scenarios where arbitrage occurs between the underlying asset market and the ETF market versus where it does not.
The results indicate that market liquidity improves in the ETF market across all metrics except Volume, while market liquidity improves in the underlying asset market specifically in terms of Volume.
Additionally, research on arbitrage between multiple assets includes studies on the transmission of declines caused by arbitrage\cite{TIY15}.
Research using artificial markets for leveraged ETFs has also been conducted\cite{ZL17}.
Regarding the impact of leveraged ETFs on the continuous trading market, Maruyama et al.\cite{YMM20a} found that volatility increases as the minimum order size for rebalancing trades decreases.
Furthermore, while it is known that an increase in the assets under management of leveraged ETFs leads to higher volatility in the leveraged ETF auction market\cite{YM16}, it remains unclear how declines in leveraged ETFs (futures) affect the futures (leveraged ETF) market.
Therefore, Mizuta et al.\cite{M24E} developed an artificial market model that included agents performing arbitrage between the leveraged ETF market (designed to be arbitrage-conscious) and the futures market. They demonstrated that when leveraged ETF (futures) prices decline, arbitrage between these markets allows leveraged ETFs (futures) to utilize futures (leveraged ETF) liquidity to limit the extent of the decline (see the Appendix for details). However, the impact of arbitrage trading on the liquidity of these markets has not been investigated.
 \par For the present study, we investigated how the liquidity of leveraged ETFs and futures markets differs when a price crash occurs in either market, depending on the presence or absence of arbitrage trading, using agent-based simulation (artificial markets).
Thus, this study extends the research of Mizuta et al.\cite{M24E}.

%\section{人工市場モデル}
\section{Artificial market model}
For the present study, we constructed a model based on the artificial market developed by Mizuta et al.\cite{M24E}\footnote{For the validity of the artificial market model used in this study, see Mizuta and Yagi\cite{MY25}.}.

\par There are two markets where futures and leveraged ETFs are traded.
Three types of agents participate in trading: normal, arbitrage, and leveraged ETF agents.
\par There are $n$ normal agents in each market, and time $t$ increases by one when a normal agent acts. There is one arbitrage agent that places orders in both markets to perform arbitrage. Finally, there is one leveraged ETF agent that places orders only in the futures market, holding futures contracts and performing rebalancing trades to maintain the leverage of the leveraged ETF.
The price determination method for each market is a continuous auction system, a system where trades are executed by matching orders in the order book (the pool of orders remaining in the market without being filled) each time an order is placed.
\par Note that this simulation does not incorporate the concepts of a day's start or end; it assumes continuous trading throughout each day.
The tick size (minimum unit of transaction price) is denoted as $\delta P$. For buy orders, fractional parts of the order price are rounded down; for sell orders, they are rounded up.
Additionally, the period up to time $t_c$ is defined as the order book formation period (described later).

%%%%%%%%

\subsection{Normal Agents}
Normal agents determine order prices by combining three strategies: the fundamental strategy, which makes investment decisions based on fundamental prices; the technical strategy, which uses past price movements to guide investment actions; and the noise strategy, which represents trial-and-error investment decisions. After determining an order price, they place buy or sell orders.
\par We here describe the order process for a normal agent, starting with the steps by which a normal agent makes a buy or sell decision.  \par The expected rate of price change $r_{e,j}^t$ at time $t$ for normal agent $j$ is calculated using equation~\eqref{re}.

\begin{equation}
\label{re}
r_e{^{t}_j}=\frac{1}{\sum_{i}^3 w_{i,j}}\left(w_{1,j}r_1{^{t}_j}+w_{2,j}r_2{^{t}_j}+w_{3,j}\epsilon{^{t}_j}\right)
\end{equation}

Here, $w_{i,j}$ is the weight for item $i$ of normal agent $j$. At the start of the simulation, the weights are initialized as uniform random numbers ranging from 0 to $w_{i,max}$.
Specifically, $w_{1,j}$ is the weight of the fundamental strategy, $w_{2,j}$ is the weight of the technical strategy, and $w_{3,j}$ is the weight of the noise strategy.
The terms of form $r_i{^{t}_j}$ are the expected returns of normal agent $j$ at time $t$.
The first term, $r_1{^{t}_j}$, represents the fundamental component return, defined by $r_1{^{t}_{j}}=\ln{\left(P_f /P^{t-1}\right)}$.
Here, $P_f$ is the constant fundamental price that does not change over time, 
and $P^t$ is the average (mid-price) of the highest bid and lowest ask prices among orders already placed at time $t$.
This definition means the term compares the fundamental price with $P^{t-1}$ predicting a positive return if the market price is lower and a negative return if it is higher.
The second term, $r_2{^{t}_j}$, represents the expected return of the technical component, defined by $r_2{^{t}_{j}}=\ln\left(P^{t-1}/P^{t-1-\tau_j}\right)$.
Here, $\tau_j$ is a uniform random number between 1 and $\tau_{max}$, determined for each agent.
Therefore, $r_2{^{t}_j}$ indicates that the agent expects a positive return if past returns were positive and a negative return if past returns were negative.
The third term, $\epsilon{^{t}_j}$, is the noise component for agent $j$ at time $t$, determined as a normally distributed random number with mean 0 and standard deviation $\sigma_\epsilon$.
Based on the predicted return derived from equation~\eqref{re}, the predicted price $P_e{^{t}_{j}}$ is determined using equation~\eqref{pe}.

\begin{equation}
\label{pe}
P_e{^{t}_{j}}=P^{t-1}exp\left(r_e{^{t}_j}\right)
\end{equation}

The order price $P_o{^{t}_{j}}$ is then determined as a uniform random number between $P_e{^{t}_{j}}-P_d$ and $P_e{^{t}_{j}}+P_d$.
If $P_o{^{t}_{j}}$ is less than $P_e{^{t}_{j}}$, the agent places a buy order with quantity 1.
If $P_o{^{t}_{j}}$ is greater than $P_e{^{t}_{j}}$, the agent places a sell order with quantity 1.
During the order book formation period, if $P_f>P_o{^{t}_{j}}$, a buy order is placed; if $P_f<P_o{^{t}_{j}}$, a sell order is placed, and $P^{t-1}=P_f$.
Furthermore, orders are canceled after $t_c$ units of time have elapsed since their placement.
\par It is generally said that leveraged ETF markets have lower liquidity than futures markets, so we set the frequency of normal agents' orders to the leveraged ETF market as probability $R_L$.

%%%%%%%%%
\subsection{Arbitrage Trading Agent}

As noted in prior research\cite{M24E}, arbitrage between actual leveraged ETFs and futures is highly complex, and determining their exchangeable price is equally intricate.
Therefore, in the model here, the arbitrage agent conducts arbitrage such that both prices satisfy equation~\eqref{plt}\footnote{For the derivation of this equation, see Mizuta et al.\cite{M24E}.}.

\begin{equation}
\label{plt}
P_L^t=P_F^t-P_{fF}\left(1-\frac{1}{L}\right)
\end{equation}

Here, $P_L^t$ and $P_F^t$ denote the market prices of the leveraged ETF and the futures contract, respectively, while $P_{fF}$ represents the fundamental price of the futures contract.
The fundamental price of the leveraged ETF is defined as $P_{fL}=P_{fF}/L$, where $L$ is the leverage ratio.
The arbitrage agent can place or cancel orders at any time.
However, this agent does not place orders during order book formation periods.
Arbitrage transactions fall into two patterns: those executed immediately and those executed after waiting until conditions are met.
These are described below.

%%%%%%%%

\subsubsection{Instant Settlement Arbitrage}
The best bid price (highest bid price in the order book) and best ask price (lowest ask price in the order book) in the futures market are denoted as $B_F$ and $S_F$, respectively. The best bid price and best ask price in the leveraged ETF market are denoted as $B_L$ and $S_L$, respectively.
Let $f(x)=x-P_{fF}\left(1-\frac{1}{L}\right)$ be the function determining how much of a leveraged ETF corresponds to the futures price $x$.
When $S_L$ is lower than the price $f(B_F)$ (which converts the price at which the futures contract can be sold into the price of the leveraged ETF), it is possible to buy the leveraged ETF cheaply and sell the futures contract at a higher price.
Therefore, the arbitrage agent immediately buys leveraged ETFs and sells futures.  In the case of $f(S_F)\leq B_L$, the opposite occurs. The arbitrage agent repeats these orders until no further orders satisfying these conditions exist.

\subsubsection{Arbitrage Transactions Executed on Hold}
Here, we first describe placing orders in the leveraged ETF market.
If $B_L<f(B_F )<S_L$, a buy order for $f(B_F)$ is placed in the leveraged ETF market.
Since this order lies between $B_L$ and $S_L$, it becomes a limit order and the trade does not execute immediately. If this order is subsequently executed, that is, if the leveraged ETF is purchased at $f(B_F)$, the futures can be immediately sold at $B_F$, enabling arbitrage at a price satisfying equation~\eqref{plt}. Similarly, even if $B_L<f(S_F )<S_L$, arbitrage can be conducted at a price satisfying equation~\eqref{plt}.
\par Next, we discuss placing orders in the futures market.
This time, since we want to determine how much of the futures price corresponds to the price $x$ of the leveraged ETF, we use the inverse function of $f(x)$: $f^{-1} (x)=x+P_{fF} (1-\frac{1}{L})$. If $B_F<f^{-1} (B_L )<S_F$, then a buy order for $f^{-1} (B_L )$ is placed on the futures contract. If this order later executes, selling the leveraged ETF at $B_L$ allows arbitrage at a price satisfying equation~\eqref{plt}.
Similarly, if $B_F<f^{-1} (S_L )<S_F$, arbitrage can be executed at a price satisfying equation~\eqref{plt}.

%%%%%%%%%

\subsection{Leveraged ETF Agent}
The leveraged ETF agent holds futures contracts.
To maintain the leverage of the leveraged ETF,
the agent conducts rebalancing trades by buying and selling the held futures.
The leveraged ETF agent decides at each time $t_R$ whether to execute a rebalancing trade.
The quantity of futures contracts the leveraged ETF agent should hold is determined by equation~\eqref{ribaransu}\footnote{For the derivation of this equation, see Mizuta et al.\cite{M24E}.}.

\begin{equation}
\label{ribaransu}
S_0\left[\frac{P_F^t}{P_{fF}} \left(L-1\right)-\left(L-2\right)\right]
\end{equation}

Here, $S_0$ is the initial futures position held by the leveraged ETF agent.
Rebalancing trades are executed when the actual position deviates by more than the ratio $w_R$ from the target position. According to equation~\eqref{ribaransu}, when the futures price $P_F^t$ rises (falls), the target futures position increases (decreases), and market buy (sell) orders are placed until the target position is reached. However, no orders are placed during the order book formation period.

\section{Simulation}

\subsection{Simulation Guidelines}
In the present study, 30 simulations were conducted under two scenarios: when erroneous orders occurred in the futures market and when they occurred in the leveraged ETF market price. Four market liquidity indicators-Volume, SellDepth, BuyDepth, and Tightness-were measured. Erroneous orders were implemented as follows.
\par The period from time $t_{ms}$ to time $t_{me}$ is defined as the erroneous order period. During this period, orders placed by each normal agent trading in the affected market are changed to market sell orders (sell orders guaranteed to execute) in the futures market with probability $p_m$.
In the leveraged ETF market, the decision to place an order occurs with probability $R_L$. If an order is placed, it is a market sell order with probability $p_m$; otherwise, a normal order is placed. Since the erroneous order is a market sell order, it executes immediately. Due to these orders being placed, prices decline during the erroneous order period.

\begin{figure}[t]
	\includegraphics[width=100mm]{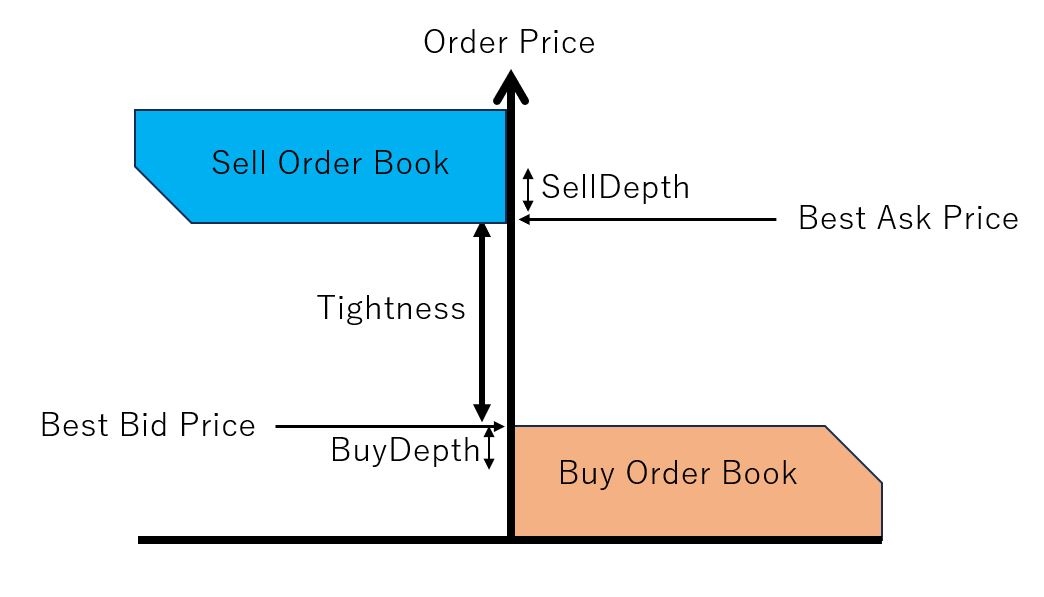}
	\caption{Market liquidity diagram\label{fig1}}
\end{figure}

This section describes  the four market liquidity indicators measured (see Figure\ref{fig1}).

Volume refers to the number of executed trades, where a higher volume indicates greater liquidity.
SellDepth refers to the thickness of the sell order book, specifically the volume of orders around the best ask price.
BuyDepth refers to the thickness of the buy order book, specifically the order volume around the best bid price. A higher order volume indicates greater liquidity.
For both depth indicators, a higher order volume means greater liquidity and the order volume was measured for 100 (50) ticks from the best bid price in the futures (leveraged ETF) market in the present study.
Tightness refers to the difference between the bid and ask prices, where a smaller difference indicates higher liquidity.
For the present study, we measured the difference between the best bid price and the best ask price (i.e., the bid--ask spread).
\par Parameters: $n=1000$, $w_{1,max}=1$, $w_{2,max}=10$, $w_{3,max}=1$, $\tau_{max}=10000$, $\sigma_\epsilon=0.03$, $P_d=P_f/10$, $t_c=20000$, $\delta P=1$, $P_{fF}=10000$, $t_{ms}=30000$, $t_{me}=60000$, $S_0=10000$, $p_m=20\%$, $R_L=20\%$, $L=2$, $t_R=10$, and $w_R=1\%$.
The simulation
period $t_e$ was set to 100,000.

%\section{実験}
%\section{結果と考察}
\subsection{Results}

\subsubsection{Consequences of Erroneous Orders in the Leveraged ETF Market}

Table~\ref{reba} shows  Volume, SellDepth, BuyDepth, and Tightness when arbitrage trading occurred versus when it did not occur under conditions where erroneous orders were placed in the leveraged ETF market.
The period covered is within the erroneous order period. The values are averages of 30 trials. Note that the bottom heading in the table indicates which market the values are for: the leveraged ETF market or the futures market. The same applies to subsequent tables.

\begin{table}[t]
  \caption{Market liquidity indicator values during the erroneous order period in the event of an erroneous order in the leveraged ETF market }
  \label{reba}
  \centering
  \begin{tabular}{|c|S|S|S|S|} \hline
%\Hline %%
  	   &\multicolumn{2}{|c|}{No arbitrage trading}&\multicolumn{2}{|c|}{Arbitrage trading permitted}\\  \cline{2-5}
 	   & {Leveraged ETF}  & {Futures} & {Leveraged ETF} & {Futures} \\  \hline
Volume &  1370.433 & 361.500 & 1336.933  &  1384.333 \\ \hline
SellDepth & 33.649 & 856.261 &  60.970 &   560.027 \\ \hline
BuyDepth & 164.257 & 855.100 &  162.713 &  934.415 \\ \hline
Tightness & 18.686 & 9.379 &   10.974 &  10.934  \\ \hline

%\Hline %%
  \end{tabular}
  
\end{table}

Table~\ref{rebabest} shows the results for the best bid and ask prices during the erroneous order period.  
\begin{table}[t]
  \caption{Best bid and best ask prices during the erroneous order period when an erroneous order occurs in the leveraged ETF market}
  \label{rebabest}
  \centering
  
  \begin{tabular}{|c|S|S|S|S|} \hline
  	   & \multicolumn{2}{|c|}{No arbitrage trading}  & \multicolumn{2}{|c|}{Arbitrage trading permitted}\\  \cline{2-5}
 	   & {Leveraged ETF} & {Futures} & {Leveraged ETF} & {Futures} \\  \hline
bestsell &4816.153&10005.410&4940.756&9940.757\\ \hline
bestbuy &4797.467&9996.033&4929.782&9929.823\\ \hline

  \end{tabular}
  
\end{table}

Table~\ref{rebatick} lists the number of orders by number of ticks  in the buy order book during the erroneous order period. Specifically, the $n$ ticks down  column indicates how many ticks below the best bid price.
For the leveraged ETF market, the depth measurement range covers the order volume from the best bid price to 50 ticks below, so values for ``100 ticks down''  are not shown.
\begin{table}[t]
  \caption{Number of orders by number of ticks  during the erroneous order period when an erroneous order occurs in the leveraged ETF market}
  \label{rebatick}
  \centering
  
  \begin{tabular}{|c|c|c|c|c|} \hline
  	   & \multicolumn{2}{|c|}{No arbitrage trading}  & \multicolumn{2}{|c|}{Arbitrage trading permitted}\\  \cline{2-5}
 	   & {Leveraged ETF} & {Futures} & {Leveraged ETF} & {Futures} \\ \hline
Best Bid Price	&2.53044&	2.532407&	1.523627&	5.494221\\ 
1 tick down	&3.107011&	3.147141&	2.883839&	7.781648\\ 
10 ticks down	&3.200591&	5.862466&	3.142820&	8.539460\\ 
20 ticks down	&3.299328&	7.956378&	3.280504&	9.025146\\ 
30 ticks down	&3.353885&	8.589546&	3.349862&	9.247427\\ 
40 ticks down	&3.394642&	8.953725&	3.424971&	9.460251\\ 
50 ticks down	&3.447853&	9.314778&	3.445988&	9.594961\\ 
100 ticks down&	 &	9.874744&	&9.828566\\

       \hline%%
  \end{tabular}
  
\end{table}

Table \ref{reba} shows that arbitrage trading did not significantly alter the Volume of the leveraged ETF market. SellDepth increased, while BuyDepth remained largely unchanged. Tightness decreased. Therefore, during the erroneous order period, the leveraged ETF market exhibited higher liquidity in terms of SellDepth and Tightness.
\par Arbitrage trading also increased Volume in the futures market.
SellDepth decreased, while BuyDepth increased.
Tightness increased. Therefore, the futures market
became more liquid in terms of Volume and BuyDepth,
but less liquid in terms of SellDepth and Tightness.
\par From the above, it can be concluded  that liquidity was supplied to either the leveraged ETF market or the futures market in terms of SellDepth and Tightness.

\subsubsection{Consequences of Erroneous Orders in the Futures Market}
Table \ref{saki} shows Volume, SellDepth, BuyDepth, and Tightness when arbitrage trading occurred versus when it did not occur under conditions where erroneous orders were placed in the futures market.
In this case, the period covered is the erroneous order period. The values are again averages of 30 trials.

\begin{table}[tb]
  \caption{Market liquidity indicator values during the erroneous order period in the event of an erroneous order in the futures market}
  \label{saki}
  \centering
  
  \begin{tabular}{|c|S|S|S|S|} \hline
 &\multicolumn{2}{|c|}{No arbitrage trading}&\multicolumn{2}{|c|}{Arbitrage trading permitted}\\  \cline{2-5}
&{Leveraged ETF} &{Futures}& {Leveraged ETF}&{Futures} \\  \hline
Volume & 68.600& 8025.533&  2508.433 &  7425.400  \\ \hline
SellDepth & 173.503 & 120.479 &  20.255 &  157.807\\ \hline
BuyDepth & 170.828 & 811.344 &  183.352 &  814.255 \\ \hline
Tightness & 5.309 & 24.343 &  17.517 &  17.594 \\ \hline

  \end{tabular}
  
\end{table}

Table~\ref{sakibest} shows the results for the best bid and ask prices during the erroneous order period.
\begin{table}[tb]
  \caption{Best bid and ask prices during the erroneous order period when an erroneous order occurs in the futures market}
  \label{sakibest}
  \centering
  
  \begin{tabular}{|c|S|S|S|S|} \hline
 %%
 %%    　     \\ \cline{2-5}
  	   & \multicolumn{2}{|c|}{No arbitrage trading}  & \multicolumn{2}{|c|}{Arbitrage trading permitted}\\  \cline{2-5}
 	   & {Leveraged ETF} & {Futures} & {Leveraged ETF} & {Futures} \\  \hline
bestsell & 5002.437  & 9552.983  &4695.787&9695.771\\ \hline
bestbuy & 4997.128&9528.640 &4678.270&9678.177\\ \hline

  \end{tabular}
  
\end{table}

Table~\ref{sakitick} lists the number of orders by number of ticks  in the buy order book during the erroneous order period. The $n$ ticks down  column indicates how many ticks below the best bid price.
For the leveraged ETF market, the depth measurement range covers the order volume from the best bid price to 50 ticks below, so values for ``100 ticks down'' are not listed.
\begin{table}[tb]
  \caption{Number of orders by number of ticks  during the erroneous order period when an erroneous order occurs in the futures market}
  \label{sakitick}
  \centering
  
  \begin{tabular}{|c|c|c|c|c|} \hline
  	   & \multicolumn{2}{|c|}{No arbitrage trading}  & \multicolumn{2}{|c|}{Arbitrage trading permitted}\\  \cline{2-5}
 	   & {Leveraged ETF} & {Futures}  & {Leveraged ETF} & {Futures}  \\  \hline
Best Bid Price	&1.812732&	4.858940&	3.190081&	4.134082\\ 
1 tick down	&1.652858&	7.717149&	3.521053&	7.626708\\ 
10 ticks down	&3.249156&	7.843822&	3.593917&	7.803561\\ 
20 ticks down	&3.494964&	7.945150&	3.679266&	7.940020\\ 
30 ticks down	&3.753843&	8.033650&	3.732040&	8.070982\\ 
40 ticks down	&3.870479&	8.116258&	3.737008&	8.163535\\ 
50 ticks down	&3.733041&	8.202681&	3.777070&	8.232720\\ 
100 ticks down&	 &	8.397247&	&8.498283\\ \hline

  \end{tabular}
  
\end{table}

Table~\ref{saki} shows that arbitrage trading increased the Volume of the leveraged ETF market.
SellDepth decreased, while BuyDepth increased.
Tightness also increased.
Therefore, the leveraged ETF market can be said to have become more liquid in terms of Volume and BuyDepth, but less liquid in terms of SellDepth and Tightness.
Arbitrage trading also caused a decrease in futures market Volume.
SellDepth increased, while BuyDepth remained largely unchanged. Tightness also decreased.
\par Therefore, from the perspective of SellDepth and Tightness, the futures market can be said to have become more liquid, but from the perspective of Volume, it can be said to have become less liquid.
\par Therefore,  from the perspective of Volume , liquidity can be considered supplied from the futures market to the leveraged ETF market.
Conversely, from the perspective of SellDepth and Tightness, liquidity can be considered supplied from the leveraged ETF market to the futures market.

%%%%%%%%%%

\section{Analysis}
\subsection{Considerations Regarding the Consequences of Erroneous Orders in the Leveraged ETF Market}
The reason leveraged ETF market Volume remained largely unchanged despite arbitrage trading is likely because, during the erroneous order period, the counterparties for normal agents' market sell orders simply shifted from other normal agents to the arbitrage agent:  the trades themselves continued under the same conditions.
This paper discusses how Tightness in the leveraged ETF market has decreased due to arbitrage trading. When the arbitrage agent places limit buy orders, these orders are executed at a price higher than the best bid, thereby setting a new best bid. 
Similarly, when limit sell orders are placed by the arbitrage agent, these orders are placed below the best ask price prior to order placement, thereby becoming the new best ask price.
Consequently, these limit orders from the arbitrage agent can be considered to have reduced the spread between the best bid and best ask prices, as shown in Table~\ref{rebabest}, thereby decreasing market Tightness.
 When a sharp decline occurs due to erroneous orders, in market models,  it is known that SellDepth increases in markets where agents such as arbitrage agent place limit buy orders\cite{YHM23}.
Normally, when the rate of decline is steep, the measurement range of SellDepth also rapidly decreases, making the sell order book invisible and causing SellDepth to appear to decrease. However, if agents like arbitrage agents continue placing limit buy orders during a sharp decline, the reduced downward price pressure makes the sell order book more visible. This, in turn, limits the decrease in SellDepth (meaning SellDepth appears to increase).
This is considered the reason why SellDepth in the leveraged ETF market increased with arbitrage trading in the present study.
The reason why the BuyDepth in the leveraged ETF market showed little change regardless of the presence of arbitrage trading is likely because, as shown in Table~\ref{rebatick}, the buy order book in the leveraged ETF market was generally thin overall. While limit buy orders placed by the arbitrage agent  caused a slight change in the number of orders near the best bid price, no difference was observed in the number of orders elsewhere on the order book. Consequently, the overall result appears unchanged.
\par The increase in futures market volume with arbitrage trading is thought to stem from the arbitrage agent placing limit buy orders in the leveraged ETF market. When these orders execute, the agent then places market sell orders in the futures market, increasing the number of executed trades and thus boosting volume.
Regarding why Tightness in the futures market increased with arbitrage trading,  the arbitrage agent placed large market sell orders in the futures market, causing declines. As shown in Table~\ref{rebabest}, while the best bid price fell, the best ask price did not decline significantly, because buy orders in the futures market remained unchanged compared to when there was no arbitrage trading.
Therefore, as shown in Table~\ref{rebabest}, the difference between the best bid and best ask prices is thought to have  increased. The reason the SellDepth in the futures market decreased when arbitrage trading occurred is thought to be because the market price fell rapidly due to market sell orders from arbitrage agents, causing the measured range of SellDepth to also drop rapidly. As a result, the sell order book became invisible, making SellDepth appear to decrease.
On the other hand, when transaction prices are stable, buying and selling occur equally, so as illustrated by the results in Table~\ref{rebatick} with no arbitrage trading, where BuyDepth only decreases near the best bid price.
However, when numerous market sell orders from the arbitrage agent occurred, the price fell, and each time, the thicker buy order book below entered the measurement range of BuyDepth.  As a result, BuyDepth in the futures market appears to have increased within its measurement range.

\subsection{Considerations Regarding the Consequences of Erroneous Orders in the Futures Market}
The decline in futures market Volume amid arbitrage activity in the present study is thought to stem from the arbitrage agent limit buy orders suppressing price declines, thereby reducing the frequency of rebalancing trades by the leveraged ETF agent. The reason arbitrage reduced Tightness in the futures market is the same as why Tightness decreased in the leveraged ETF market when erroneous orders occurred earlier: the bid--ask spread narrowed due to limit orders to buy and sell placed by the arbitrage agent (see Table\ref{sakibest}).
The increase in SellDepth in the futures market due to arbitrage trading is for the same reason that SellDepth increased in the leveraged ETF market when erroneous orders occurred in that market.
This is because if agents persistently place limit buy orders like the arbitrage agent during a sharp decline, the downward price pressure weakens, making the sell order book more visible.  Consequently, the decrease in SellDepth is also suppressed.
On the other hand, the reason BuyDepth remained largely unchanged is essentially the same as why it showed little variation when erroneous orders occurred in leveraged ETFs: as shown in Table~\ref{sakitick}, while limit buy orders from the arbitrage agent slightly altered the number of orders near the best bid price, no differences were observed in the order book beyond that. Consequently, the overall result appears  unchanged.
\par The increase in leveraged ETF market Volume with arbitrage activity
is thought to stem from the arbitrage agent placing market sell orders in the leveraged ETF market when their limit buy orders in the futures market are executed.
Regarding the increase in Tightness in the leveraged ETF market with arbitrage, while the best bid price declined as the arbitrage agent placed large market sell orders in the leveraged ETF market, causing a price drop (as shown in Table \ref{sakibest}), the best ask price did not decline significantly, because buy orders in the leveraged ETF market remained unchanged compared to when there was no arbitrage.
Therefore, as shown in Table \ref{sakibest}, the spread between the best bid and ask prices is thought to have  widened and increased.
The reason the SellDepth in the leveraged ETF market decreased due to arbitrage trading
is the same reason why SellDepth in the futures market decreased when arbitrage trading was present during erroneous orders in the leveraged ETF market. It is because market sell orders from the arbitrage agent accelerated the price decline, causing the measured range of SellDepth to rapidly decrease as well. Consequently, the sell order book became invisible, making SellDepth appear to decrease.
On the other hand, the reason BuyDepth increased is that when erroneous orders occur in the leveraged ETF market, the futures market's BuyDepth behaves similarly to when arbitrage trading exists. Since trading prices are stable, buying and selling occur equally. Therefore, as shown in Table~\ref{sakitick} (without arbitrage trading), the BuyDepth in the leveraged ETF market only causes the values near the best bid price to decrease.
However, when a large number of market sell orders from the arbitrage trading agent occur,  prices decline, and each time this happens, the lower buy order book enters the measurement range of BuyDepth.
As a result, BuyDepth in the leveraged ETF market appears to have increased within the measurement range.

\section{Conclusion}
The present study investigated the difference in market liquidity between leveraged ETFs and futures markets when a price crash occurs in either market, using artificial market simulations to examine the impact of arbitrage trading. This study extended the research by Mizuta et al.\cite{M24E}.
\par The investigation revealed that when erroneous orders occur in the leveraged ETF market, the presence of arbitrage trading causes liquidity to flow from the futures market to the leveraged ETF market in terms of SellDepth and Tightness. When erroneous orders occur in the futures market, the existence of arbitrage trading causes liquidity to flow from the leveraged ETF market to the futures market in terms of SellDepth and Tightness, and from the futures market to the leveraged market in terms of Volume.
Furthermore, by investigating the internal mechanisms of the artificial market, we were able to elucidate the reasons behind these results.
\par However, since the present study does not incorporate the concept of a trading day, it could not address cases where rebalancing transactions are concentrated at the end of the day. This should be considered a topic for future research.

\section*{Funding Statement}
This work was supported by JSPS KAKENHI [grant number 23K04276].
The authors thank FORTE Science Communications (https://www.forte-science.co.jp/) for English language editing.
\section*{Declarations}
It should be noted that the opinions contained herein are solely those
of the authors and do not necessarily reflect those of SPARX Asset
Management Co., Ltd.

\bibliography{btxj28}
\bibliographystyle{plain}

\appendix
\section{Research results by Mizuta et al.\cite{M24E}}
\begin{figure}[t]
	\includegraphics[width=100mm]{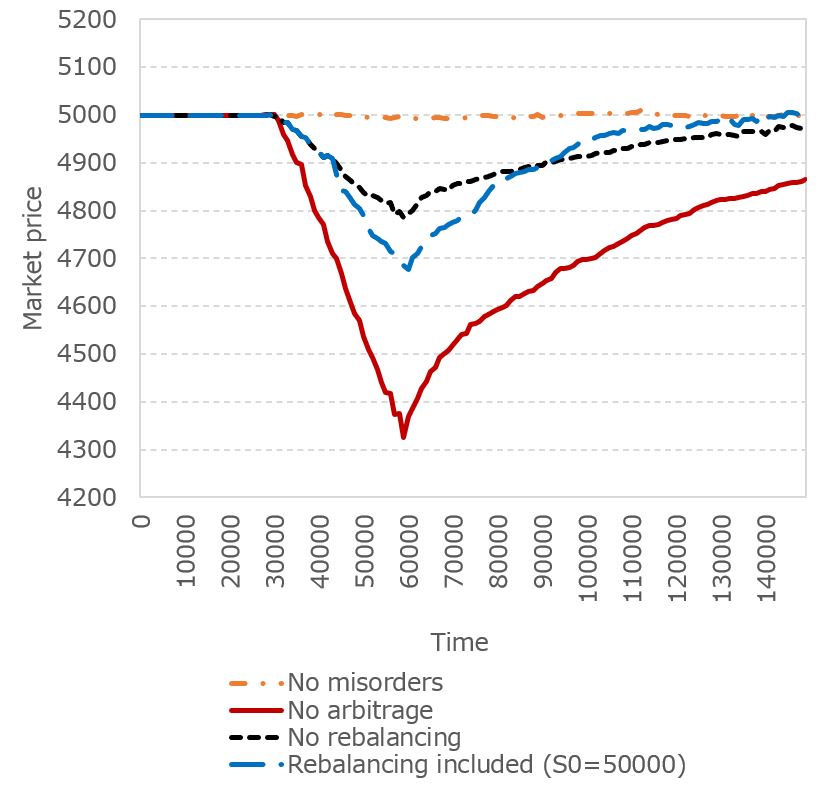}
	\caption{Price time series under the following conditions: no erroneous orders or arbitrage transactions, erroneous orders present but no arbitrage transactions, erroneous orders and arbitrage transactions present but no rebalancing transactions, or erroneous orders, arbitrage transactions, and rebalancing transactions present\label{e01}}
\end{figure}

The research results of Mizuta et al.\cite{M24E} are described below.
 Figure~\ref{e01} shows the price time series of leveraged ETFs under four scenarios: no erroneous orders or arbitrage trading; erroneous orders but no arbitrage trading; erroneous orders and arbitrage trading but no rebalancing trades; and erroneous orders, arbitrage trading, and rebalancing trades ($S_0 = 50000$).
It can be seen that the price drops significantly when erroneous orders occur.
 Furthermore, it is understood that arbitrage trading can limit declines. Normal agents' sell orders for leveraged ETFs match with arbitrage agents' buy orders. Since arbitrage agents place sell orders for futures, it is the buy orders for futures that are consumed. The decline is mitigated thanks to arbitrage agents' buy orders for leveraged ETFs, and they can place these orders precisely because they can consume the buy orders for futures. In other words, the decline in leveraged ETFs
can be mitigated using the liquidity of futures contracts.
This fact  indicates that futures liquidity serves as hidden liquidity for leveraged ETFs.
On the other hand, it is also evident that rebalancing transactions can amplify the magnitude of declines.

\begin{figure}[t]
	\includegraphics[width=100mm]{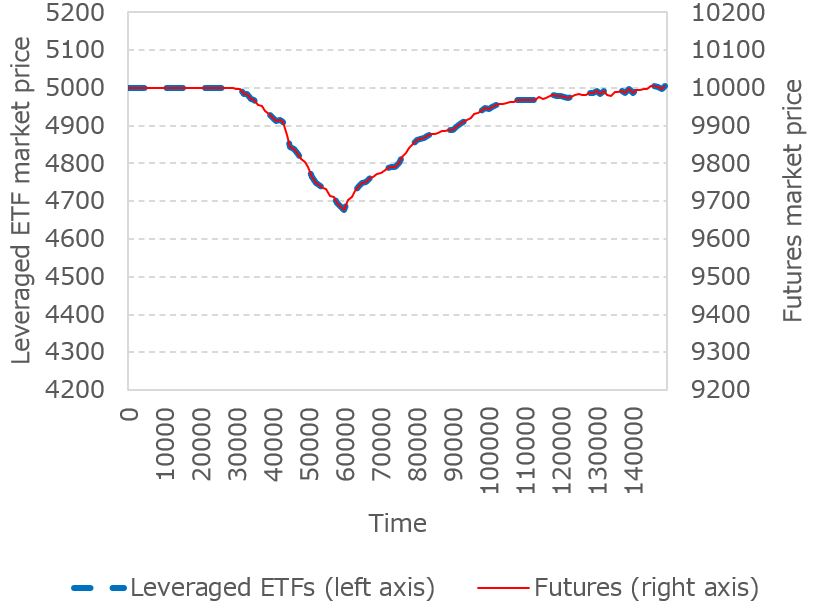}
	\caption{Price movements of leveraged ETFs and futures with misorders, arbitrage, and rebalancing ($S_0=50000$). Although both the left and right vertical axes span a $1000$ range, the initial price of the leveraged ETF is $5000$, half that of the futures, at $10000$. Therefore, a $2$-fold change in the leveraged ETF will cause the graphs to overlap.\label{e02}}
\end{figure}

\begin{figure}[t]
	\includegraphics[width=100mm]{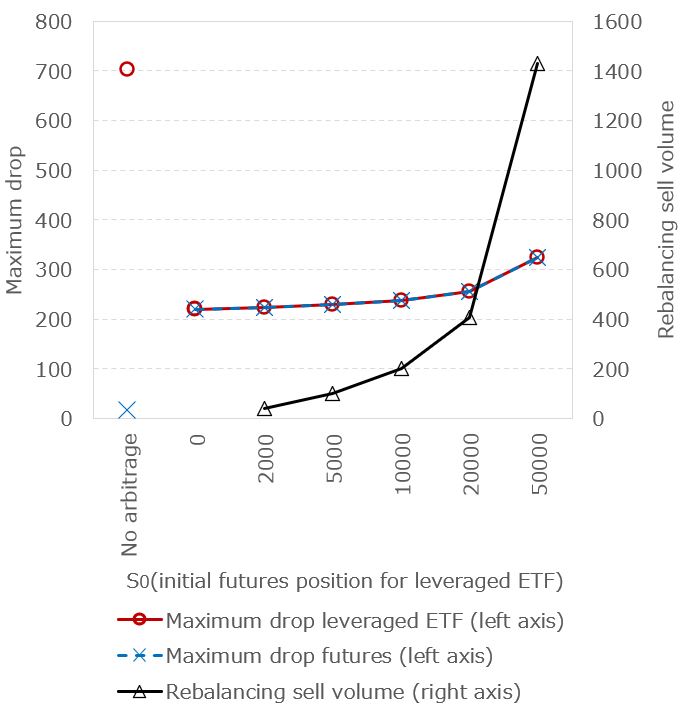}
	\caption{Incorrect orders placed, no arbitrage, and the decline in leveraged ETFs when $S_0$ is altered with arbitrage, along with the sell volume resulting from leveraged ETF rebalancing.\label{e03}}
\end{figure}

\begin{figure}[t]
	\includegraphics[width=100mm]{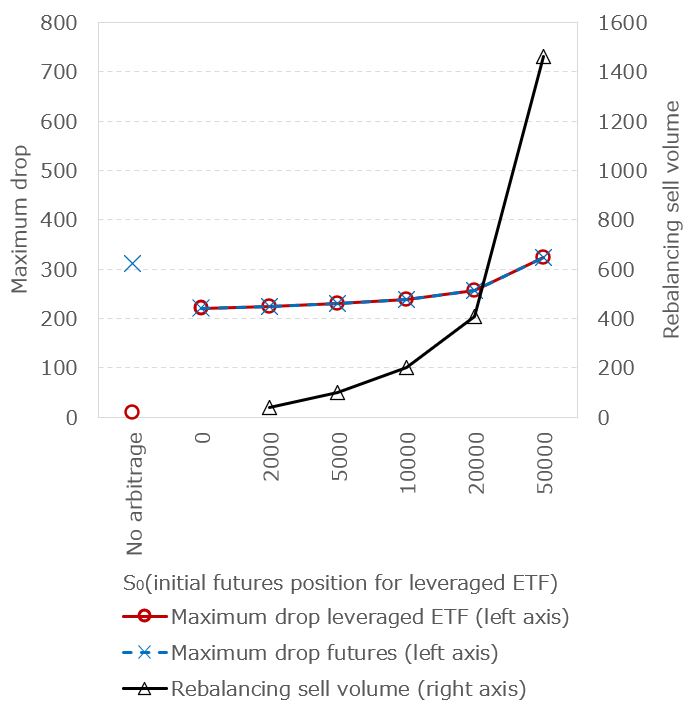}
	\caption{Incorrect orders placed, no arbitrage, and the decline in futures when $S_0$ is altered with arbitrage, along with the sell volume resulting from leveraged ETF rebalancing.\label{e04}}
\end{figure}

\begin{figure}[t]
	\includegraphics[width=100mm]{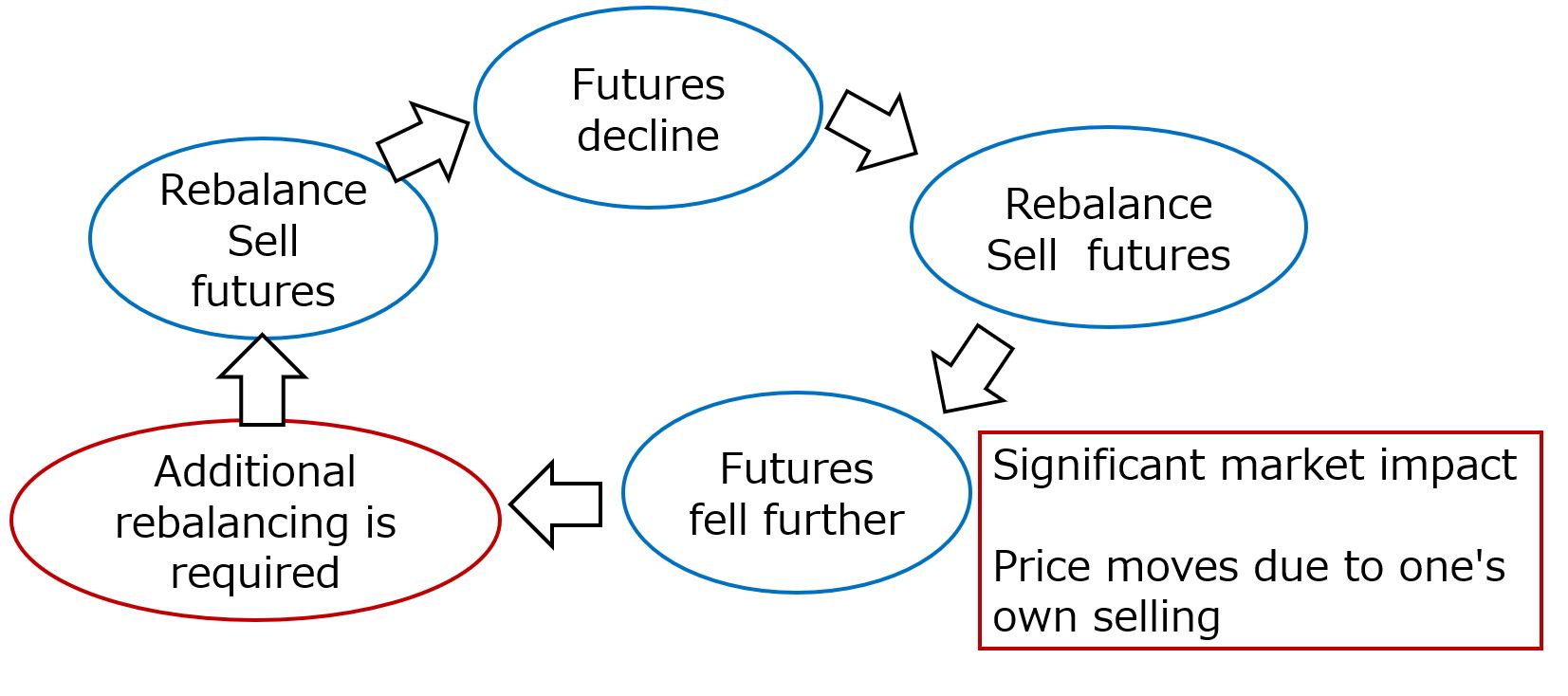}
	\caption{Mechanism of market disruption when leveraged ETFs become too large ($S_0 \ge 100000$)\label{p03}}
\end{figure}

 Figure~\ref{e02} shows the price movements of leveraged ETFs and futures when mispricing, arbitrage, and rebalancing are present
($S_0=50000$). Substituting the parameters  defined in this Appendix(L=2,$P_{fF}=10000$)  into equation \eqref{plt} yields $P^t_L=P^t_F-5000$, and we can see that arbitrage achieves this result.
Note that although both the left and right vertical axes span a $1000$ range, the initial price of the leveraged ETF is $5000$, half that of the futures contract, at $10000$. Therefore, the graph will align when the leveraged ETF experiences a $2x$ fluctuation.
 Next, with incorrect orders present and no arbitrage, and with arbitrage present,
$S0 = 0, 2000, 5000, 10000, 20000, 50000$, while keeping all other conditions identical, including the random number table, the decline amount ($P_f - minimum price$) and the sell quantity due to leveraged ETF rebalancing were calculated.
This process was repeated $100$ times by changing the random number table, and the average value was used.

 Figure\ref{e03} shows the case of erroneous orders in leveraged ETFs. In the absence of arbitrage trading, only leveraged ETFs experience a significant decline, while futures do not fall.
When arbitrage trading exists, as mentioned earlier, arbitrage allows leveraged ETFs to utilize futures liquidity, reducing the magnitude of their decline. However, as leveraged ETFs grow larger (as $S_0$ increases), the magnitude of their decline increases.
Furthermore, for $S_0 \ge 100000$, the decline was too large for the simulation to run.

 Figure~\ref{e04} shows a case where an erroneous order was placed in futures.
Thanks to arbitrage trading, futures can utilize the liquidity of leveraged ETFs, reducing the magnitude of the decline. However, the decline in the absence of arbitrage trading is not as severe as in Figure \ref{e03}. This is because leveraged ETFs have low order volume $(w_R=20\%)$ and are prone to declines due to erroneous orders, whereas futures with high order volume are less susceptible to declines caused by erroneous orders.

 However, when arbitrage trading occurs, the result is nearly identical to Figure~\ref{e03}. Arbitrage trading combines the liquidity of both instruments, leading to a similar reaction to erroneous orders.  From the futures perspective, the liquidity received from the less liquid leveraged ETFs is relatively small, resulting in a smaller dampening effect on declines. From the perspective of investors trading leveraged ETFs, the benefits of high futures liquidity are significant. However, from the perspective of investors trading futures, the benefits derived from the liquidity of leveraged ETFs are not substantial.

Indeed, for $S_0 \ge 100000$, the decline was too large for the simulation to run. This can be interpreted as representing market disruption.
Figure~\ref{p03} illustrates the mechanism of market disruption in this case, specifically when leveraged ETFs become excessively large.
When futures decline, leveraged ETFs sell futures.
When leveraged ETFs are large (holding a significant amount of futures), the quantity of futures to sell also increases according to equation~\eqref{ribaransu}.
This selling then causes the futures price to fall, leading to further declines in the futures,
requiring additional rebalancing sales.
This creates a loop that cannot be stopped, causing the market decline to persist and resulting in market disruption.

Yagi et al.\cite{YMM20a,YM16}  demonstrated that when rebalancing causes market impact greater than volatility, the resulting price fluctuations necessitate additional rebalancing, which in turn requires further rebalancing, creating an unstoppable loop. This is consistent with that finding.

\end{document}